\documentclass{article}
\usepackage{arxiv}

\usepackage{graphicx}

\usepackage[appendix=inline]{apxproof}

\usepackage{comment}

\usepackage[utf8]{inputenc} 
\usepackage[T1]{fontenc}    
\usepackage{hyperref}       
\usepackage{url}            
\usepackage{booktabs}       
\usepackage{amsfonts}       
\usepackage{nicefrac}       
\usepackage{microtype}      
\usepackage{lipsum}		
\usepackage{graphicx}
\usepackage{doi}
\usepackage{amssymb}
\usepackage{amsmath}

\newtheorem{example}{Example}
\newtheorem{definition}{Definition}

\usepackage{amssymb}

\newcommand{\A}{\approx}
\newcommand{\B}{\not\approx}

\newcommand{\D}{\overset{?}{>}_G}
\newcommand{\M}{\overset{?}{>}_G}

\newcommand{\gc}{G_C}

\usepackage{amsmath,amssymb}
\usepackage{array}
\usepackage{xcolor}
\usepackage{soul}
\usepackage{enumitem}
\usepackage{float}

\usepackage{todonotes}
\newtheoremrep{theorem}{Theorem}
\newtheoremrep{lemma}{Lemma}

\usepackage{amsmath,amssymb}
\usepackage{array}

\title{Dynamic E-unification}

\author{ \href{https://orcid.org/0009-0009-3846-5771}{\includegraphics[scale=0.06]{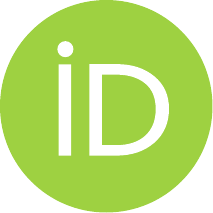}\hspace{1mm}Kun Han} \\
	Clarkson University\\
    8 Clarkson Avenue\\
	Potsdam, NY 13699-5815 \\
	\texttt{hank@clarkson.edu} \\
	\And
	\href{https://orcid.org/0000-0003-1141-0665}{\includegraphics[scale=0.06]{orcid.pdf}\hspace{1mm}Christopher Lynch} \\
	Clarkson University\\
    8 Clarkson Avenue\\
	Potsdam, NY 13699-5815 \\
	\texttt{clynch@clarkson.edu} \\
}

\date{}

\begin{document}
\maketitle




\begin{abstract} 

We present an E-unification 
procedure for a set of non-ground (dis)equations, along with a dynamic set of ground (dis)equations, and prove its completeness.  The ground part is dynamic in the sense that it continually changes.  The algorithm saturates the non-ground equations using Superposition modulo the ground theory.  We also have an Instantiation rule that matches the left hand side of non-ground (dis)equations with ground terms, creating new ground (dis)equations, which changes the ground
theory. This algorithm can be used in quantified SMT problems, where the dynamic ground theory represents the evolving model.   We develop an ordering to compare terms modulo a ground theory, which is used to orient non-ground equations.  We prove properties of this ordering, using a weak form of monotonicity and subterm property.
We finally present a set of inference rules for our ordering, which allows us to properly orient equations in theories of some finite data structures, such as a theory of finite lists with length and append.         

\end{abstract} 
\section{Introduction}



Satisfiability Modulo Theories has emerged as an efficient method for deciding satisfiability of large formulas~\cite{DBLP:series/faia/BarrettSST21}.  The formulas are represented by large sets of ground clauses.  Satisfiability is normally decided modulo a theory.  Some theories are built-in, while other theories can be defined using small sets of non-ground clauses.  We focus on theories defined by a set of non-ground (dis)equations.  The propositional part of the SMT solver, along with a congruence closure procedure, will construct a set of (dis)equations which serves as a model of the ground clauses.   At the high level, the SMT solver tries to decide the satisfiability of that ground model in the non-ground theory.  This is an equational unification problem~\cite{DBLP:books/el/RV01/BaaderS01}, since it proves unsatisfiability by finding instances that make the two sides of an instance of a ground disequation equal.  We call this a dynamic E-unification problem, because when the ground model is unsatisfiable in the theory, the SMT solver can come up with a new model.  We want an E-unification algorithm that works efficiently when the model changes.

The way that SMT solvers deal with this problem is to select a set of triggers for each (dis)equation in the non-ground theory.  A trigger is a subterm of the (dis)equality.  All variables must be included in the set of triggers.  The SMT solver will find a substitution that matches ground terms onto the triggers, modulo the ground theory.  This process is called E-matching~\cite{DBLP:conf/cade/MouraB07}.  
The substitution is used to instantiate the (dis)equation,
and a new ground (dis)equation is created.  This process does not always halt.  If it does halt and the triggers are chosen well, the correct instances will be created to show unsatisfiability when the problem is unsatisfiable.  If the process halts on a satisfiable problem, the SMT solver will not know that it is satisfiable.   So SMT solvers can often prove unsatisfiability but not satisfiability.  Some SMT solvers are able to build a finite model for some satisfiable cases~\cite{DBLP:conf/cav/GeM09}, but that often does not work.

Alternatively, rewriting techniques are used for E-unification.  Knuth-Bendix completion is applied to the equations~\cite{KnuthBendix1970}, and narrowing~\cite{hullot1980canonical} is applied to the disequations.  They must be applied to both non-ground and ground (dis)equations.  Since we are assuming a large number of ground equations, the completion procedure creates too many equations.  Another option is to only perform completion on non-ground equations, but to perform the inferences modulo the equational theory of the ground equations.  This would require lots of extensions to the non-ground equations.
Additionally, it requires an ordering compatible with the ground theory, which is difficult to achieve.   Another difficulty using rewriting techniques is that we want a dynamic method that allows the ground theory to change.

To address these difficulties, we develop a saturation-based SMT method.  We saturate the non-ground (dis)equalities ($NG$) modulo the ground theory ($G$).  We must also orient the $NG$ equations modulo $G$.   We don't know in advance what the final $G$ will be, so we develop an ordering ($G$-ordering) to compare terms modulo any $G$.
%
Our first task is to create a well-founded reduction ordering that is compatible with the ground theory.  It is well-known that this is a difficult problem.  We compare terms by comparing their normal forms under a confluent rewrite system representing $G$, even though we don't know what that is.
Unfortunately, this destroys monotonicity and the subterm property.  However, we define weaker properties, which we call non-theory monotonicity and the non-theory subterm property, which are good enough for our purposes.  

We also
instantiate the non-ground terms, as an SMT solver does.  The instantiation procedure is performed when an instance of the entire left hand side of an $NG$ equation is equivalent modulo $G$ to a member of $G$.  This is related to the extension procedure in completion modulo an equational theory, but since the instantiated equation is ground, it is added to $G$ instead of $NG$, effectively changing $G$.  As opposed to the SMT procedure, we also need to perform superposition modulo the equational theory, similar to completion modulo an equational theory.
If we saturate under our inference rules, and the ground theory is satisfiable, then we are guaranteed that the entire set of clauses is satisfiable.  

This process seems to work well with constructor theories, such as the theory of lists with length and append, as shown here:

\begin{example}
\label{LLA}
\emph{LLA} is the theory of lists with the functions 
length and append.

\begin{flushleft}
\(
\text{(1)}\ {car}({cons}(X, Y)) \A X
\)\\
\(
\text{(2)}\ {cdr}({cons}(X, Y)) \A Y
\)\\
\(\text{(3)}\
{len}({cons}(X, Y)) \A s({len}(Y))
\)\\
\(\text{(4)}\
{app}({cons}(X, Y), Z) \A {cons}(X, {app}(Y, Z))
\)\\
\(\text{(5)}\
{app}({nil}, W) \A W
\)\\
\(\text{(6)}\
{s}( X) \B zero 
\)\\
\(\text{(7)}\
{len}( {nil}) \A zero \)
\end{flushleft}
\end{example}

All LLA (dis)equations are oriented under our $G$-ordering, except for Equations 3 and 4. 
If our saturation procedure, run under this orientation, results in SAT, then we run an inference procedure to detect a cycle in the list definitions.  
If the set of equations represents finite lists, then we will not detect a cycle, and we show that this set of dis(equations) is satisfiable in the theory of finite lists.

The paper is organized as follows.  We start with preliminaries, where we prove important properties of congruence classes.  Next we define an ordering modulo $G$ and prove this ordering has the required properties.  We then define a saturation procedure using this ordering.  
Finally, we give a procedure to allow a weaker orientation procedure, and use that to show how to show satisfiability in the LLA theory.  We conclude by discussing related and future work.  

\section{Preliminaries}
We follow standard definitions for rewriting~\cite{dershowitz1990rewrite}.
We assume we are given a set of variables, which we represent with capital letters, 
and a
set of uninterpreted function symbols of various arities, represented with lower case letters.  An
arity is a non-negative integer.  {\em Terms} are defined
recursively in the following way:  each variable is
a term, and if $t_1,\cdots,t_n$ are terms, and $f$
is of arity $n \geq 0$, then $f(t_1,\cdots,t_n)$ is a term.
If $s$ and $t$ are terms then $s \A t$ is an equation and $s \B t$ is a disequation.
An object is {\em ground} if it has no variables, otherwise {\em non-ground}.

Let $G$ be a set of ground (dis)equations, with a 
congruence relation generated by the equations of $G$, written $=_G$. 
For each ground term $t$, we denote a constant $Rep(t)$ as the chosen representative
of the congruence class of $t$. 
We assume an
ordering $<$, with $Rep(t)$ the smallest term in the congruence class of $t$.
%
Assume that $G$ is presented as a terminating and confluent
ground rewrite system 
whose rewrite rules are compatible with a reduction
ordering $<$.  
For any ground term $t$, we denote by $t\!\downarrow_G$ the unique normal form
of $t$ under rewriting by $G$. 
Therefore, for ground terms $s$ and $t$, 
$s =_G t$ iff $s\!\downarrow_G = t\!\downarrow_G$.
\begin{example}

Let
\(
G = \{\; cons(a,b) \approx c,\; len(c) \approx d \;\}.
\)  The congruence classes are $\{ \{a\}, \{b\},
\{c, cons(a,b)\}, \{d, len(c)\} \}$, and the rewrite system is 
$\{cons(a,b) \rightarrow c, len(c) \rightarrow d\}$.
%
%
\end{example}

A term $t$ is called a subterm of another term $s$, written as $t \sqsubseteq s$, if:
    (1) $t=s$ or 
    (2) $s$ has the form $f(t_1, ..., t_n)$, and $t \sqsubseteq t_i$ for some $t_i$.
Let \(Sub(G)\) denote the set of all subterms of terms
occurring in \(G\).
The set of subterms modulo \(G\) is: 
\[
   Sub_G(G)
   \;=\;
   \{\, t \mid \exists \; s \in Sub(G)\ \text{with}\ s =_G t \,\}.\footnote{Note that $ t\!\downarrow_G=Rep(t)$ if $t \in Sub_G(G).$}
\]


A {\em substitution} is a mapping from the set of 
variables to the set of terms, which is 
almost everywhere the identity.  
We identify a substitution with its homomorphic
extension.  Composition of substitutions $\sigma$ and $\rho$ is defined so that $X\sigma\rho = (X\sigma)\rho$ for all variables $X$.
A substitution $\theta$ matches $A$ to $B$ if $A\theta = B$, and
is a {\em unifier} of
$A$ and $B$, if $A\theta = B\theta$.
$\sigma$ is a {\em most general unifier} of $A$ and $B$, written
$\sigma = mgu(A,B)$ if $\sigma$ is a unifier of $A$ and $B$,
and for all unifiers $\theta$ of $A$ and $B$, there is a 
substitution $\rho$ such that $X\sigma\rho = X\theta$ for all $X$ in $Vars(A \cup B)$.
A substitution $\sigma$ is a \emph{solution} to $E$ if
\(
s\sigma =_G t\sigma
\;\text{for all } s = t \in E.
\)
%
A set of substitutions $\Theta \in CSU_G(E)$ if
(1) all members of $\Theta$ are solutions of $E$ and 
    (2) if $\theta$ is a solution of $E$, then there exists $\sigma \in \Theta$ and a substitution $\tau$ such that
    \(
        E\sigma\tau = E\theta.
    \)





We give a few lemmas involving congruence classes that will be useful in this paper.  For these lemmas, we need to formally define a proof in $G$.

\begin{definition}
A proof that $s =_G t$ is a finite sequence of terms
\(
s_0 = s_1 = s_2= \ldots = s_n
\)
where
\(
s = s_0, \; t = s_n,
\)
and for all $i < n$ there exists a ground equation $u = v \in G$ with
\(
s_i = s_i[u] \;\text{and} \; s_{i+1} = s_i[v].
\)
For any $i$ with $s_i = u$ and $s_{i+1} = v$, we call this a step at the top.
\end{definition}

Membership in $Sub_G(G)$ is preserved under subterm.

\begin{lemmarep}
\label{sub}
If \(s[t] \in Sub_G(G)\), then \(t \in Sub_G(G)\).
\end{lemmarep}

\begin{proof}
We prove the contrapositive.
Assume that \(t \notin Sub_G(G)\).
We show that \(s[t] \notin Sub_G(G)\).

Suppose, toward contradiction, that \(s[t] \in Sub_G(G)\).
Then there exists some \(s' \in Sub(G)\) such that
\(
   s[t] =_G s'.
\)
Consider a proof of
\(s[t] =_G s'\).

\medskip\noindent
{Case 1:
No step of the proof occurs at or above the occurrence of \(t\). } 

Then there exists \(t' = _Gt\) such that \(t'\) is a subterm of \(s'\),
with \(t' \in Sub(G)\), so \(t \in Sub_G(G)\).

\medskip\noindent
{Case 2:
There is a step at or above the position of \(t\).}

Consider the first such step. Then the step has the form
\(
   s'[t'] = _G  u \; \text{where}\; t' = _G t.
\)
In particular, \(t' \in Sub(G)\).
But then \(t \in Sub_G(G)\), again contradicting the assumption.

\medskip

In both cases, we derive a contradiction, so our assumption that
\(s[t] \in Sub_G(G)\) while \(t \notin Sub_G(G)\) is impossible.
Hence, the contrapositive holds, and the lemma follows.
\end{proof}

For every $f$-term in $Sub_G(G)$, there is an equivalent $f$-term in $Sub(G)$.

\begin{lemmarep} \label{f_in_sgg}
If $f(s_1,\dots,s_n) \in {Sub}_G(G)$, there exists $f(t_1,\dots,t_n) \in {Sub}(G)$ such that $s_i =_G t_i$ for all $i$.
\end{lemmarep}
\begin{proof}
Let $f(s_1,\!\dots,\!s_n) \in {Sub}_G(G)$, and $t \in {Sub}(G)$ such that $f(s_1,\!\dots,\!s_n) =_G t$. 

\medskip
\noindent{Case 1: There is no step on the top of the proof of  $ f(s_1, \ldots, s_n) =_G t$.}  
Since there is no step on the top of the proof, \(t\) must be of the form  $f(t_1,\dots,t_n)$ with  \( s_i =_G t_i \) for all \( i \).

\medskip
\noindent{Case 2: There is a step on the top of the proof of \(  f(s_1, \dots, s_n) =_G t \).}  
Assume the first step of the proof at the top is
\(
f(t_1, \dots,  t_n) =_G v,
\)
where \( f(t_1, \dots,  t_n) = v \) is in \( G \).

Since there is no previous step on the top of the proof, \( s_i =_G t_i \) for all \( i \).
Since \( f(t_1, \ldots, t_n) = v \in G \), it follows that \( f(t_1, \ldots, t_n) \in {Sub}(G) \).
Therefore, there exists \( f(t_1, \ldots, t_n) \in {Sub}(G) \) such that \( s_i =_G t_i \) for all \( i \).

\medskip
In both cases, we obtain the required $f(t_1,\dots,t_n) \in {Sub}(G)$ with $s_i =_G t_i$ for all $i$.  
\end{proof}

If a term is equivalent to a direct subterm of itself then both are in $Sub_G(G)$.

\begin{lemmarep} \label{f_t_in_sgg}
If
\(
f(t_1,\dots,t_n) =_G t
\), and \(t \sqsubseteq t_i\), then 
$
f(t_1,\dots,t_n)$ $\in {Sub}_G(G).
$
\end{lemmarep}
\begin{proof}
Assume, for contradiction, that there exists a term \( f(t_1, \ldots, t_n) \)
such that \( f(t_1, \ldots, t_n)=_G t \), \(t\) is a subterm of some \(t_i\), and \( f(t_1, \ldots, t_n) \notin {Sub}_G(G) \).
Consider the smallest such term with respect to the size of \( f(t_1, \ldots, t_n) \).

Since  \( f(t_1, \ldots, t_n) \notin {Sub}_G(G) \), there can be no step at the top of \( f(t_1, \ldots, t_n)=_G t \), so let \( t = f(s_1, \ldots, s_n) \). Since \( t \sqsubseteq t_i \), We can write 
\( f(\ldots, t_i[t], \ldots)=_G f(s_1, \ldots, s_n) \), 
where for each \( j \), \( t_j =_G s_j \), in particular, \( t_i[t] = s_i \). 

By the minimality of our counterexample, \( t_i  \in {Sub}_G(G) \) implies \( t  \in {Sub}_G(G) \) by Lemma~\ref{sub}. So \( f(t_1, \ldots, t_n) \in {Sub}(G) \). Contradiction.
\end{proof}

Subterms not in $Sub_G(G)$ do not lose their structure under $G$-equivalence.

\begin{lemmarep}\label{subset_lemma}
Let $G$ be a set of ground equations.  
If $u' =_{G} u[s]$ and $s \notin Sub_{G}(G)$, then there exists $s' \sqsubseteq u'$ such that
\(
s' =_{G} s .
\)
\end{lemmarep}

\begin{proof}
Since $u' =_G u[s]$, there exists a finite $G$-derivation
\(
u[s] = s_0 \;\leftrightarrow_G\; s_1 \;\leftrightarrow_G\; \cdots \;\leftrightarrow_G\; s_n = u',
\)
where each step rewrites a subterm using an equation from $G$. Consider the position $p$ of the occurrence of $s$ in $u[s]$.
Because $s \notin Sub_G(G)$, no  step in the derivation can occur
at position $p$ or above it.

Therefore, all steps occur below position $p$.
It follows that the subterm at position $p$ is preserved throughout the
derivation up to $G$-equivalence.
Let $s' = u'|_p$ , then
\(
s' =_G s.
\)

By construction, $s' \sqsubseteq u'$, and we have $s' =_G s$, as required.
\end{proof}

\section{The Ordering \texorpdfstring{$<_G$}{<G}}

We assume that $<$ is a well-founded reduction ordering (well-founded, and monotonic)
that is total on ground terms and compatible with rewriting (if $t \rightarrow s$ under the rewriting system, then $s < t$ in the reduction ordering).
We define a new order $<_G$, based on a set of ground equations $G$.  The idea is that we want a compatible ordering modulo the ground theory.

\begin{definition}[The ordering $<_G$]\label{<g} 
    $s <_{G} t$ iff
    $s \!\downarrow_G < t\!\downarrow_G$.

\   
\begin{example}
Let $G = \{ cons(a,b) = nil,\;len(nil) = zero,\; s(c) = zero,\; len(b) = c \}$ and the ground
constant ordering is $c \succ b \succ nil \succ a\succ zero$.
Then
\[
len(b)\!\downarrow_G =  c,
\quad
len(cons(a,b))\!\downarrow_G =  len(nil)\!\downarrow_G=zero.
\]
Since $c \succ zero$, we have $len(b)\!\downarrow_G > len(cons(a,b))\!\downarrow_G$, and therefore
$len(b) >_G len(cons(a,b))$.
\end{example}

\end{definition}


We next show that $<_G$ is an ordering and it has some nice properties.
The first three properties follow directly from the properties of $<$.

    \begin{theoremrep} [Irreflexivity]   
    For any term $t$, it is not the case that $t <_{G} t$.
    \end{theoremrep}
    \begin{proof}

    If we assume $t<_{G} t$, then $t\!\downarrow_G < t\!\downarrow_G$, which is impossible, by reflexivity of $<$. Thus, $<_G$ is irreflexive. 
    \end{proof}

    \begin{theoremrep}[Transitivity]
    If $s<_{G} t$ and $t<_{G} u$, then $s<_{G} u$. 
    \end{theoremrep}

    \begin{proof} 
     
    Since $s <_{G} t$ gives $s\!\downarrow_G < t\!\downarrow_G$ and $t<_{G} u$ gives $t\!\downarrow_G < u\!\downarrow_G$, we conclude $s\!\downarrow_G < u\!\downarrow_G$ by transitivity of $<$. So $s <_{G} u$.
     \end{proof}

    \begin{theoremrep}[Well-Foundedness]    
    $<_G$ is a well-founded ordering on ground terms. 
    \end{theoremrep}
    \begin{proof}
    Since $<$ is well-founded, there is no infinite descending chain
    ${t_{0}\!\downarrow_G} > {t_{1}\!\downarrow_G} > \cdots$, i.e., there is no infinite descending chain $t_0 >_G t_1 >_G \cdots$.   
\end{proof}

Unfortunately, the $<_G$ ordering is not monotonic, and does not obey the subterm property.  For example, suppose that $G = \{
f(a) \approx c\}$, where $f(a) > f(b) > f(c) > a > b > c$.  Then $a >_G b$ but $f(b) >_G f(a)$, and $a >_G f(a)$.  So we define a weaker form of monotonicity and subterm property.  In particular, the properties are restored if the terms are not in $Sub_G(G)$.
   
\begin{theoremrep}[Nontheory Monotonicity]
    If $s >_{G} t$, and $f(s_1,\cdots, s_{i-1}, s,s_{i+1},\cdots, s_n) \notin Sub_G(G)$ then  $f(s_1,\cdots, s_{i-1}, s, s_{i+1},\cdots, s_n) >_{G}f(s_1,\cdots,s_{i-1}, t, s_{i+1},\cdots, s_n)$.    
\end{theoremrep}
\begin{proof}
  
   Since $f(s_1, ..., s_n) \notin Sub_G(G)$, we have: 

            ${f(s_1, ..., s_{i-1}, s, s_{i+1}, ..., s_n) \!\downarrow_G}$ $= $ ${f({s_1\!\downarrow_G}, ..., {s_{i-1}\!\downarrow_G},{s\!\downarrow_G},{s_{i+1}\!\downarrow_G},,..., {s_n\!\downarrow_G})}$.

            Since $s >_{G} t$, and by monotonicity of $>$, we can get that\\
\(
f({s_1\!\downarrow_G}, ..., {s_{i-1}\!\downarrow_G},{s\!\downarrow_G},{s_{i+1}\!\downarrow_G},,..., {s_n\!\downarrow_G})
>
f({s_1\!\downarrow_G}, ..., {s_{i-1}\!\downarrow_G},{t\!\downarrow_G},{s_{i+1}\!\downarrow_G},\)\\
\(...,{s_n\!\downarrow_G})\geq
  f({s_1\!\downarrow_G}, ..., {s_{i-1}\!\downarrow_G},{t\!\downarrow_G},{s_{i+1}\!\downarrow_G},..., {s_n\!\downarrow_G})\!\downarrow_G = f(s_1, ..., s_{i-1}, t, s_{i+1}, ..., s_n)\!\downarrow_G\).

Therefore, \(
f(s_1, ..., s_{i-1}, s, s_{i+1}, ..., s_n) \!\downarrow_G
>
f(s_1, ..., s_{i-1}, t, s_{i+1}, ..., s_n) \!\downarrow_G
\).

Hence $f(s_1, ..., s_{i-1}, s, s_{i+1}, ..., s_n) >_{G}f(s_1, ..., s_{i-1}, t, s_{i+1}, ..., s_n)$.
\end{proof}

\begin{theoremrep}[Nontheory subterm property]
If \(s \notin Sub_G(G)\) and \(t\) is a proper subterm of \(s\), then
\(
s >_G t.
\)

\end{theoremrep}

\begin{proof}
We prove this property by contradiction.

Suppose the claim is false: there exist ground terms
\(s,t\) with \(t\) a proper subterm of \(s\), \(s\notin{Sub}_G(G)\), and yet
\(
\neg(s >_G t).
\)
By totality, from \(\neg(s >_G t)\) we must have either \(s =_G t\) or \(t >_G s\).  The equality case \(s =_G t\) is
ruled out by the hypothesis \(t\) is a proper subterm of \(s\) and \(s\notin{Sub}_G(G)\), hence we must have
\(
t >_G s.
\)

Write $s$ as $s[t]$.
Using nontheory monotonicity and Lemma~\ref{sub}, we obtain
\(
t >_G s = s[t] \Longrightarrow s[t] >_G s[s[t]].
\)
Applying the same nontheory monotonicity step again yields
\(
s[t] >_G s[s[t]] >_G s[s[s[t]]] >_G \cdots .
\)
Chaining with the initial \(t >_G s=s[t]\) we obtain an infinite descending chain
\(
t >_G s[t] >_G s[s[t]] >_G \cdots,
\)
which contradicts the well-foundedness.

Therefore the assumption \(\neg(s>_G t)\) is impossible; hence \(s >_G t\).
\end{proof}

The last properties are fairly straightforward from the properties of $<$.

\begin{theoremrep}[Totality on $G$-Classes] For all $s$, $t$: $s <_{G} t$, $s >_{G} t$ or $s =_{G} t$. 
\end{theoremrep}
\begin{proof}

We analyze the relationship between $s$ and $t$ as follows:

\begin{enumerate}
   \item If $s\!\downarrow_G < t\!\downarrow_G,$ then by Definition~\ref{<g}, $s <_{G} t$  
    \item If $s\!\downarrow_G > t\!\downarrow_G,$ then by Definition~\ref{<g}, $s >_{G} t$ 
    \item If $s\!\downarrow_G = t\!\downarrow_G,$ then $s =_{G} t$.
    \end{enumerate}

\end{proof}

\begin{theoremrep}[$G$-Compatibility] If $s' =_G s >_{G} t =_{G} t'$, then:  $s' > _{G} t'$.     
\end{theoremrep}  
\begin{proof}
We have $s' \!\downarrow_G = s \!\downarrow_G $, and $t \!\downarrow_G = t' \!\downarrow_G$.
Therefore, 
   $s'\!\downarrow_G = s\!\downarrow_G > t\!\downarrow_G = t'\!\downarrow_G$.
    So, $s' > _{G} t'$.
    \end{proof}

Equations are compared using the multiset ordering, with the equation considered as a pair.  Disequations are compared the same way.  Disequations are always larger than equations.  Non-ground equations and disequations are compared. In general, $s >_G t$ if $s\sigma >_G t\sigma$ for all ground substitutions $\sigma$.

%
  
\section{The Sup(\texorpdfstring{$G$}{G}) Inference System}


Our inference system $Sup(G)$ works on a set of clauses $S$, which is the disjoint union of $NG$ (non-ground clauses) and $G_C$ (ground clauses).
The set $NG$ consists of  Horn clauses, with only one 
(dis)equation.  
There may not be ground disequations initially in 
$NG$, but Superposition inference rules add them. 
$G$ is a set of equations and disequations, which is a model of $G_C$.    

The $Sup(G)$ inference rules are performed on clauses from $NG$ and are given in the table above.  The $\bowtie$ in the rules is either $\approx$ or $\not\approx$.  The results of the inference rules are added back to $S$.  For Instantiation, the conclusion is guaranteed to be ground, because if there is no larger side with respect to $>$ then both sides are instantiated.
In the Superposition rule, we unify using a substitution $\sigma$ where $\sigma$ is a member of the complete set of unifiers of $s$ and $s'$
modulo $G$(for example using the algorithm from \cite{DBLP:conf/tacas/BarbosaFR17}), 
and $\Delta  \subseteq G$, is a set of equations where \( s \sigma =_\Delta t \sigma\). 
$\neg\Delta$
is added to the conclusion, to preserve soundness,
because modifying $G_C$ may modify $G$.

We will prove that the saturation of an unsatisfiable set of clauses will produce an unsatisfiable ground theory.  As usual, Simplification and Subsumption are optional rules.  Simplification replaces the premise with the conclusion, and Subsumption deletes the premise.




\subsection{The Sup(\texorpdfstring{$G$}{G}) Inference Rules}

\begin{figure}[H]
\centering
\begin{tabular}{|c|c|c|}
\hline
\textbf{Rule Name} & \textbf{Inference Rule} & \textbf{Side Conditions} \\
\hline

\text{Instantiation}
&
\(
\dfrac{\neg\Gamma \lor s \,\A\, t}{\neg\Gamma \lor (s \,\A\, t)\sigma}
\)
&
\begin{minipage}{5.8cm}
\vspace{0.5mm}
\centering
\begin{itemize}
  \item $s\sigma \in {Sub}_G(G)$ if $t \ngeq  s$
  \item $t\sigma \in {Sub}_G(G)$ if $s \ngeq t$
  \item $G \models \Gamma$
\vspace{0.5mm}
\end{itemize}

\end{minipage}
\\
\hline

\text{Superposition}
&
\(
\dfrac{\neg\Gamma_1 \lor s \A t \;\; \neg\Gamma_2 \lor u[s'] \,\!\bowtie\!\, v}{\neg \Delta \!\lor\! \neg\Gamma_1 \!\lor\! \neg\Gamma_2 \!\lor\!( u[t] \,\!\bowtie\!\, v)\sigma}
\)
&


\begin{minipage}{6.2cm}
\vspace{0.5mm}
\centering
\begin{itemize}
  \item \(G \models  \Gamma_1 \) and
 \(G \models  \Gamma_2 \)
  \item $t  \ngeq _G s $
  and $v \ngeq _G u[s']$
  \item $s'$ is not a variable
  \item $s\sigma \notin {Sub}_G(G)$
  \item $\sigma \in {CSU_G}(s, s')$
  \item \(\Delta \subseteq G\) is a conjunction of ground equations such that \(\Delta \models s \sigma = t \sigma\)

  \vspace{0.5mm}
\end{itemize}

\end{minipage}

\\
\hline

\shortstack{Equality \\Resolution}
&
\(
\dfrac{\neg \Gamma \lor u \B v}{\neg\Delta  \lor \neg \Gamma \lor (u \B v) \sigma} 
\)
&
\begin{minipage}{5.8cm}
\vspace{0.5mm}
\begin{itemize}[topsep=0pt, partopsep=0pt, parsep=0pt, itemsep=2pt]
    \item \(G \models  \Gamma \) 
  \item $\sigma \in {CSU_G}(u, v)$
  \item \(\Delta \subseteq G\) is a conjunction such that \(\Delta \models u \sigma = v \sigma\)
      \vspace{0.5mm}
\end{itemize}

\end{minipage}
\\
\hline

\text{Simplification}
&
\(
\dfrac{
\neg\Gamma \lor u[s']\,\bowtie\,v
}{
 \neg\Gamma \lor u[t\sigma]\,\bowtie\,v
}
\)
&
\begin{minipage}{5.8cm}
\vspace{0.5mm}
\centering
\begin{itemize}
 \item  $s \A t \in NG $ 
  \item  $s\sigma > _G t\sigma$ and
  \((u[s']\,\bowtie\,v) > _G(s \A t)\) for all $G$\footnotemark
    \item $s\sigma = s'$
  \vspace{0.5mm}
\end{itemize}

\end{minipage}

\\
\hline

\text{Subsumption}
&
Remove clause $D$
&
\begin{minipage}{5.8cm}
\vspace{0.5mm}
\centering
$\exists$ 
    \( C \in NG \)    such that $C\sigma \subseteq D$
   \vspace{0.5mm}
\end{minipage}
\\
\hline
\end{tabular}
\caption{The Sup($G$) Inference Rules}
\end{figure}
\footnotetext{For example, if $t \sqsubset s$}


\begin{example}
  {Sup($G$) example}
\label{supg}

We assume that
$G = \gc = \{ len(nil) \A zero, cons(a,b) \A nil \}$.
\\[5pt]
\(
\dfrac{app(cons(X,Y),Z) \A cons(X, app(Y,Z)) \quad app(nil,W) \A W}{cons(a,b)\B nil \;\lor\; cons(a,app(b,W))  \A W} \quad\text{Superposition}
\)\\[5pt]
\(
\dfrac{len(cons(X,Y)) \A s(len(Y)) }{len(cons(a,b)) \A s(len(b))}\quad\text{Instantiation }
\)
\\[5pt]
\(len(cons(a,b)) \A s(len(b))\) is added to \(\gc\)
.\\[5pt]
\(
\dfrac{ s(X) \B zero}{s(len(b)) \B zero}\quad\text{Equality Resolution}
\)
\\[5pt]
\(s(len(b)) \B zero\) is added to  \(\gc\) 
.\\[5pt]
Now 
\(\gc\) 
is unsatisfiable, since 
\(s(len(b)) =_G len(cons(a,b)) =_G len(nil)=_G zero\), which conflicts with \(s(len(b)) \B zero\).
\end{example}

\subsection{Completeness of Sup(\texorpdfstring{$G$}{G}) Inferences Rules}


Next we prove completeness of Sup($G$), meaning that for an unsatisfiable set of clauses, we are guaranteed to produce an unsatisfiable ground set of clauses.    As usual, we consider a saturated set up to redundancy.  Superposition, Instantiation and Equation Resolution are required for completeness.  Simplification and Subsumption remove redundant clauses.  Simplification has been defined so that it does not rely on the ground part. We use a model construction proof, which is similar to the usual model construction, but we create the model modulo $G$.

Let's examine the differences between our proof and the traditional model construction proof.  We start with a model $G$ of the ground clauses, and then build a model of the non-ground equations $NG$, modulo $G$.  We consider an instance $\neg\Gamma \vee s \bowtie t$ only if the equations of $\Gamma$ are true in $G$, and we reduce the instance by $G$.  We add an instance to the model if the LHS is irreducible by smaller equations in terms of $<_G$, but we require in addition that the LHS is not in $Sub_G(G)$.  The reason for this is that, for non-variable positions, we will need to apply Superposition to this instance to get a smaller counterexample.  The result of Superposition will be smaller by nontheory monotonicity of $<_G$, since the LHS is not in $Sub_G(G)$.  If the LHS is reducible at a variable position, then Lemma~\ref{subset_lemma} implies that we can reduce the substitution.
On the other hand, if the LHS is in $Sub_G(G)$ then Instantiation can be applied, showing that this (dis)equation was already true in $G$.



\begin{itemize}
  \item \textbf{Redundancy} A clause $C$ is redundant in $NG$ if, for every ground substitution $\theta$,  
  the instance $C\theta$ is implied by smaller ground instances of $NG$.
  \item \textbf{Saturation} $NG$ is $\text{Sup}(G)$ saturated if the conclusion of every $\text{Sup}(G)$ inference from $NG$ is either subsumed by $NG$ or redundant in $NG$.
\end{itemize}   


\begin{definition}
    We are given \( G \), a model of the ground clauses \( \gc \). 
A model is a confluent and terminating set of ground equations. 
We treat $M$ as if $s \B t \in M$ if and only if $s \ne_M t$.  In other words, we implicitly assume that $M \models s \B t$ if and only if $M \not\models s \A t$.

Let\ \(
{Gr}_G(NG) = \left\{ (s\sigma)\!\downarrow_G \A (t\sigma)\!\downarrow_G \;\middle|\; \neg\Gamma \lor s \A t \in NG,\ \sigma \text{ is ground,} \;G \models \Gamma \right\}\)
\end{definition}

\begin{definition}
We define $M_G^{s \A t}$ and $M^{<s \A t}_G$ co-recursively.
\( M_G^{s\A t} = \{ s \A t \} \) if:

\begin{enumerate}
  \item \( s >_G t \) and
  \item \( s \notin {Sub}_G(G) \) 
  \item \( s \) is not reducible by \( M_G^{<s\A t} \)
\end{enumerate}

Otherwise, \( M_G^{s\A t} = \emptyset \).

We define:
$
M^{<s \A t}_G = \bigcup \{ M^{u \A v}_G \mid \ (u \A v) <_G (s \A t), u \A v \in Gr_G(NG) \}$


Let $
{M^\infty_G}= \bigcup \{ {M^ {s\A t}_G} \mid s \A t \in {Gr_G}({NG}) \}, and \  M =  G \cup {M^\infty_G}\}
$\\
\end{definition}


\begin{theoremrep}
If $NG$ is saturated by Sup($G$), and \( G \) is a model of the ground clauses \( \gc \), then $M \models {Gr_G}({NG}) \cup G$. 

\end{theoremrep}

\begin{proof}
We suppose that $M \not\models {Gr_G}({NG}) \cup G$, 
and then there exists $u \,\bowtie\, v \in {Gr_G}({NG}) \cup G$ 
such that $M \not\models u \,\bowtie\, v$. Let $u \,\bowtie\, v$ be the smallest element of ${Gr_G}({NG}) \cup G$ 
such that $M \not\models u \,\bowtie\, v$ .
We may assume without loss of generality that $u >_G v$.
We consider the following cases:
\subsection*{Case 1: Equations}

In this case, \(u \,\bowtie\, v \) is of the form \( u \A v \).

We know that $u \A v$ is not in $G$.
Since $u \A v \in {Gr_G}(NG)$, there exists a non-ground
clause
$\neg \Gamma_1 \lor u' \A v' \in NG$ and a ground substitution $\sigma$ such that
\( G \models \Gamma_1, 
u = u'\sigma \!\downarrow_G, \; v = v'\sigma
\!\downarrow_G \).

\begin{enumerate}
 
  \item \textbf{$u \in {Sub}_G(G)$:} 
Since $u \in Sub_G(G)$ and $u >_G v$ then $v \in Sub_G(G)$.
So Instantiation can be applied, resulting in 
$\neg \Gamma_1 \vee u'\sigma \A v'\sigma$ is in $\gc$,
because $NG$ is saturated. Contradiction.  So $M \models u \A v$.

\item {$u \notin {Sub}_G(G)$ and $u$ is reducible by \( M_G^{<u\A v} \):}
Suppose \( u = u[s] \), and there exists an equation \( s \A t \in {Gr}_G(NG) \) such that
\( u[s] \A v \in {Gr}_G(NG) \).

There exists
\( \neg \Gamma_2 \lor s'' \A t'' \in NG \) such that:
\(G \models \Gamma_2, \;
s''\sigma \!\downarrow_G = s, 
\; t''\sigma \!\downarrow_G = t
\).
Let \( \Delta \subseteq G\) such that 
$s =_{\Delta} s''\sigma$, $t =_{\Delta} t''\sigma$,
$u =_{\Delta} u'\sigma$ and $v =_{\Delta} v'\sigma$.

By Lemma~\ref{subset_lemma}, one of the two following cases must hold:

\begin{enumerate}
  \item 
  \( u' \) is of the form \( u'[s'] \), where \( s' \) is not a variable and 
  \( s'\sigma \!\downarrow_G = s \). 
  Then there is an inference of the following form:
  \[
  \frac{\neg \Gamma_1 \lor u'[s'] \A v' , \quad \neg \Gamma_2 \lor s'' \A t''}{\neg \Delta \lor \neg\Gamma_1 \lor \neg\Gamma_2 \lor u'[t''] \theta \A v' \theta}
  \]
  where \( u'[t'']\sigma \!\downarrow_G = u[t] \) and \( v'\sigma \!\downarrow_G = v \).
  Hence, \( u[t] \A v \) is an instance of \( u'[t''] \theta \A v' \theta \).

  Since \(u[t] \A v <_G u[s] \A v\),
the equation \(u[t] \A v\) is a smaller counterexample by nontheory monotonicity if it exists.  Otherwise, it is redundant, which also yields a smaller counterexample.
  
  \item 
   \( u' \) is not of the form \( u'[s'] \), where \( s' \) is not a variable and 
  \( s'\sigma \!\downarrow_G = s \), then there exists \(x\sigma = w[s']\)  
  such that \( s'\!\downarrow_G = s \) due to \( s \notin {Sub}_G(G) \).
 
  Let \( \sigma' \) be a substitution such that
  \[
  y\sigma' = 
  \begin{cases}
  y\sigma & \text{if } y \neq x,\\[4pt]
  w[t] & \text{if } y = x.
  \end{cases}
  \]
  by nontheory monotonicity.
  Then \( u'\sigma'\!\downarrow_G  \A v'\sigma'\!\downarrow_G \) is a smaller counterexample, contradicting the minimality of \( u \A v \).
\end{enumerate}

\item 
 \text{\( u \notin {Sub}_G(G) \) and \( u \) is not reducible by  \( M_G^{<u\A v} \):}

Then \( M_G^{u\A v} = \{ u \A v \} \).

Thus, \( M \models u \A v \).
\end{enumerate}

In all cases, we reach a contradiction. Hence, 
\(
M \models {Gr_G}({NG}).
\)

\subsection*{Case 2: Disequations}

In this case, \(u \,\bowtie\, v \) is of the form \( u \B v \).
%
%
    Suppose for contradiction that  \(u \B v\ \in G\)
    is \emph{false} in \(M\), i.e.
\(
M \models u \A v.
\)  
Assume both sides are normalized with respect to \(G\).
%
%
The disequation is either from $G$
or $Gr_G(NG)$.
  
\begin{enumerate}
    \item $u \B v$ is in $G$.
\begin{enumerate}
    \item   \( u = v \). \\
    This case is impossible.
    \item   \( u \ne v \). \\
Because \(M \models u \A v\), there is an \(M\)-rewrite proof which reduces \(u\). Let the first step that reduces \(u\): \(
u = u[s] \xrightarrow{s \A t} u[t],
\)
for some ground equation $s \A t $.

We lift this ground step as follows:
there exist $\neg \Gamma_1 \lor s' \A t' \in NG$ and a ground substitution $\sigma$ such that
\( G\models \Gamma_1, \;
s'\sigma \!\downarrow_G = s,
\; \text{and} \;
t'\sigma \!\downarrow_G = t
\).

Since \(u \B v \in G\)
and \(u \in {Sub}_G(G)\), it follows that \(s' \sigma \in {Sub}_G(G)\), therefore $t'\sigma \in Sub_G(G)$.
By Instantiation,
the ground instance $s'\sigma \A t'\sigma$ is in $G$,
since $NG$ is saturated.  This leads to a contradiction, because \( u \A v \) is assumed to be normalized and therefore cannot be further reduced.
\end{enumerate}

\item $u \B v$ is in \(Gr_G(NG)\)

 Since $u \B v \in {Gr_G}(NG)$, there exists a non-ground
clause
 $\neg \Gamma_1 \lor u' \B v' \in NG$ and a ground substitution $\sigma$ such that:
\( G \models \Gamma_1, \;
u = u'\sigma \!\downarrow_G, \; v = v'\sigma
\!\downarrow_G\).
Let \( \Delta \subseteq G\) such that 
$u =_{\Delta} u'\sigma$ and $v =_{\Delta} v'\sigma$.

\begin{enumerate}

  \item
   $u = v$. 
   
   Then there is an Equality Resolution inference:

  \[
  \frac{\neg \Gamma_1 \lor u' \B v' }
  {\neg\Delta  \lor \neg \Gamma_1 \lor (u \B v) \sigma} 
  \]

  But then $\neg \Gamma_1$ is in $\gc$, which is a contradiction, since $G \models \Gamma_1$.

  \item
  $u \ne v$.



Then \( u = u[s] \), and there exists an equation \( s \A t \in {Gr}_G(NG) \) such that
\( u[s] \A v \in {Gr}_G(NG) \).  Also, 
there exists
\( \neg \Gamma_2 \lor s'' \A t'' \in NG \) such that:
\(G \models \Gamma_2, \;
s''\sigma \!\downarrow_G = s, 
\; t''\sigma \!\downarrow_G = t.
\)
Let \( \Delta \subseteq G\) such that 
$s =_{\Delta} s''\sigma$, 
$t =_{\Delta} t''\sigma$, 
$u =_{\Delta} u'\sigma$ and
$v =_{\Delta} v'\sigma$, 

We know \(s \notin Sub_G(G)\).
So, by Lemma~\ref{subset_lemma}, \(u'\sigma = u' \sigma[s'] \; \text{where} \; s'=_Gs\).
  \begin{enumerate}
\item   \( u' \) is not of the form \( u'[s'] \), where \( s' \) is not a variable and 
  \( s'\sigma \!\downarrow_G = s \), then there exists \(x\sigma = w[s']\)  
  such that \( s'\!\downarrow_G = s \) due to \( s \notin {Sub}_G(G) \).
 
  Let \( \sigma' \) be a substitution such that
  \[
  y\sigma' = 
  \begin{cases}
  y\sigma & \text{if } y \neq x,\\[4pt]
  w[t] & \text{if } y = x.
  \end{cases}
  \]
  by nontheory monotonicity. Then \( u'\sigma'\!\downarrow_G \A v'\sigma'\!\downarrow_G \) is a smaller counterexample, contradicting the minimality of \( u \A v \).

  \item
\( u' \) is of the form \( u'[s'] \), where \( s' \) is not a variable and 
  \( s'\sigma \!\downarrow_G = s \). 
  Then there is an inference of the following form:
 
  \[
  \frac{\neg \Gamma_1 \lor u'[s'] \B v' , \quad  \neg \Gamma_2 \lor s'' \A t''}{\neg\Delta \lor \neg\Gamma_1 \lor \neg\Gamma_2 \lor u'[t''] \theta \B v' \theta}
  \]
  where \( u'[t'']\sigma \!\downarrow_G = u[t] \) and \( v'\sigma \!\downarrow_G = v \).
  Hence, \( u[t] \B v \) is an instance of \( u'[t''] \theta \B v' \theta \).

  Since \(u[t] \B v\) is smaller than \(u \B v\) with respect to \(>_G\),
then either the equation \(u \B v\) is a smaller counterexample by nontheory monotonicity, or it was removed by redundancy and there is some smaller counterexample.
  \end{enumerate}
\end{enumerate}
\end{enumerate}
\end{proof}

\section{Ordering for finite data structures}


We have shown the completeness of Sup($G$) if the equations are oriented under $<_G$.  Since $<_G$ is weaker than $G$, we sometimes have to orient equations in both directions.
In this section modify the lexicographic path ordering to use the ground theory to detect that the 
ground equations represent finite objects, for example finite lists.  First we saturate the equations under the ordering $<$.
After saturation, we check an ordering on the saturated set.  If we don't find a cycle\footnote{A cycle is a chain of constants $c_0 > c_1 > \cdots c_n$ where $c_0 = c_n$.}  in the equations, then 
there is an ordering on the constants so that the $<_G$ ordering matches the $<$ ordering.  

\subsection{Lexicographic Path Ordering}
We recall the definition of LPO.


\begin{definition}
{Lexicographic Path Ordering (\(>_{{LPO}}\)).}  

Let \(\prec\) be a precedence on function symbols.  
The \emph{lexicographic path ordering} \(>_{{LPO}}\) is the smallest transitive relation satisfying the following conditions:

Given terms $s$ and $t$, \(s > _{{LPO}} t\) if \(\;t \sqsubset s\), otherwise let \(s = f(s_1, \ldots, s_n)\) and \(t = g(t_1, \ldots, t_m)\):
\begin{enumerate}
  \item  \(g \prec f\) and \(s > _{{LPO}} t_i\) for all \(i\);
  \item  \(f = g\), and  there exists \(i\), such that
        \(
          s_i >_{{LPO}} t_i,
          \text{and if} \; j < i, \text{then} \;
           s_j = t_j, 
          \text{ and if } \; j > i,  \text{then} \; s
          > _{{LPO}} t_j; 
         \)
 \item  \(f \prec g\), and there exists i, such that \(s_i >_{{LPO}} t\) for some \(i\).
\end{enumerate}
\end{definition}

\subsection{The LPOGC Inference System}

We modify the LPO ordering so that it can detect the existence of a cycle in the equations.  When it does not find a cycle, we know that there is an ordering on the constants so that the given orientation of the equations is sufficient.


First we define a function called $gc$, which returns a set of ordering constraints on constants.  The purpose of $gc$ is to instantiate non-ground ordering constraints.  It will be used in our inference system to find all necessary instantiations, in order to determine an ordering on the ground constants.  The $gc$ function has two arguments.  The first one is an ordering constraint that we want to instantiate, and the second argument is a boolean, whose value is 0 if all instantiations are included and 1 if none of them are.

\begin{definition}

\(gc(f(s_1,\dots,s_n)>t,b) = \{\, f(t_1,\dots,t_n)\!\downarrow_G > t\theta\!\downarrow_G\ \mid f(t_1,\dots,t_n)\in {Sub}(G), \forall \hspace{0.25em} 1\le i \le n , \hspace{0.25em} s_i\theta =_G t_i, \text{ and } \, b = 0\} \).
%
\end{definition}

\begin{example}
\(G=\{\,cons(a,c)=c\}.\)
\[
gc(cons(X,Y) > Y, 0)
=
\{\, cons(a,c)\!\downarrow_G > y\theta\!\downarrow_G \,\}
=
\{\, c > c\,\}.
\]

\end{example}

Now we define the LPOGC Inference System, whose inference rules operate on a pair of components, separated by a semicolon.  The first component is a set of
pairs, each pair contains an ordering constraint and a boolean value. We sometimes abbreviate a set $\{(s_1 \D t_1, b), \cdots (s_n \D t_n, b)\}$ as $\{s_1 \D t_1, \cdots s_n \D t_n\}_b$.
The second component is a set of necessary constraints on the ground constants.  Whenever $b = 0$, constraints are added to this set.  Initially, $b = 1$, because we are assuming the initial left hand side is not in $Sub_G(G)$, and nothing needs to be added in that case.


\textbf{The LPOGC Inference System}

Let
\(s = f(s_1, \ldots, s_n)\) and \(t = g(t_1, \ldots, t_m)\): 

\begin{enumerate}
    \item lpog:
\[
\frac{\{\, s \D t \}_b\, \cup \Delta; \;C}{\{\,s \D t_1,\;\dots,\;s \D t_m\;\}_b\, \cup \Delta; \;C \; \cup\; gc( s> t,b)}
\]
where \(g \prec f\)

    \item lpoe:
    \[
\frac{\{\, s \D t \}_b\, \cup \Delta; \;C}{\{ \; s_i \D t_i\;\}_0\;\cup\; \{\, s \D t_{i+1}, \; \dots, \; s \D t_m\;\}_b\, \cup \Delta;  \;C \; \cup\; gc( s> t,b)}
\]
where 
\begin{itemize}
    \item \(f = g\)
    \item \(\,s_1 = _Gt_1,\;\dots,\;s_{i-1} =_G t_{i-1}\)
    \item \(1\le i \le n\)
\end{itemize}
  
    \item lpol:
    \[
\frac{\{\, s \D t \}_b\, \cup \Delta; \;C}{\{\,s_i \D t \}_0\, \cup \Delta;  \;C \; \cup\; gc( s> t,b)}
\]
where
\begin{itemize}
    \item \(f \prec g\)
\item \(1\le i \le n\)
\end{itemize}
\item lpos:
 \[
\frac{\{\, s \D t \}_b\, \cup \Delta; \;C}{ \Delta;  \;C \; \cup\; gc( s> t,b)}
\]
where \(t\sqsubset s\)
\end{enumerate}

Now we define the LPOGC Inference System, whose inference rules operate on a pair of components, written $S;C$.  $S$ is a set of
pairs, containing an ordering constraint and a boolean value. A set $\{(s_1 \D t_1, b), \cdots (s_n \D t_n, b)\}$ is written as $\{s_1 \D t_1, \cdots s_n \D t_n\}_b$.
$C$ is a set of necessary constraints on the ground constants.  Whenever $b = 0$, constraints are added to this set.  Initially, $b = 1$, because we are assuming the initial left hand side is not in $Sub_G(G)$, and nothing needs to be added in that case.

The proof is by induction.  If the first component is $\emptyset$ and $C$ does not contain a cycle, then the $C$ gives a well-founded ordering which can extend the well-founded ordering on the theory symbols.  Otherwise, we assume that this extended ordering satisfies the bottom of an inference, and show that it must also satisfy the top.  Let $(s >_G t, b)$ be the constraint on the top of the inference.  If $s\theta \not\in Sub_G(G)$ then $s\theta >_G t\theta$ by the definition of LPO.  If $s\theta \in Sub_G(G)$, then we can see that $b = 0$, by the rules of LPOGC.  So the inference adds the proper constraints to make $s\theta >_G t\theta$.

\begin{theoremrep}
    
Let  > be an LPO order, and
\(
U=\{u_1,\dots,u_n\}
\) be a finite set of non-ground terms, and \(\theta\) be a ground substitution
such that, for all $i$, \(u_i \theta \notin {Sub}_G(G)\).  
If we can derive $\emptyset ; C$ from 
\(
\,\{\,u_1 \M v_1,\; \dots,\; u_n \M v_n\,\}_1;\;\emptyset
\),
where \(C\) does not contain a cycle, then there exists an ordering on the symbols \emph{consistent with \(C\)} such that LPO satisfies
\(
u_1\theta >_G v_1\theta,\;\dots,\; u_n\theta >_G v_n\theta.
\)
\end{theoremrep}

\begin{proof}

Extend the precedence to the constants so that the precedence on constants is consistent with the constraint set $C$.  

We prove the statement by induction on the number of steps in the derivation leading
to $\emptyset; C$.  

Let $s \M t$ be a constraint occurring in the derivation of $\emptyset; C$. Assume that all constraints  on the bottom are satisfied by $\theta$; we must then show
that the constraints on the top are satisfied by $\theta$.

\begin{enumerate}
    \item Suppose  \(s\theta\notin Sub_G(G)\)

Let \(
s = f(s_1,\dots,s_n)\).

We now prove that \(s >_G t\) by a case analysis on the
function symbols \(f\) and \(g\).
    \begin{enumerate}
        \item \(g \prec f\), and let \(t = g(t_1,\dots,t_m)\).
        

\[
\frac{\{\, s \D t \}_b\, \cup \Delta; \;C}{\{\,s \D t_1,\;\dots,\;s \D t_m\;\}_1\, \cup \Delta; \;C \; \cup\; gc( s> t,b)}
\]
We assume that \(
s_1>_G t_1,\; \dots\; s_n>_G t_m
\), and then 
\(
s_i\theta\!\downarrow_G > t_i\theta\!\downarrow_G\,(1\le i\le m).
\)
We can get that:
\(
s\theta\!\downarrow_G =\! 
f(s_1\!\downarrow_G,\! \dots,\! s_n\!\downarrow_G)
\ >\ 
g(t_1\!\downarrow_G,\dots,t_m\!\downarrow_G)
\ge g(t_1, \dots, t_m)\!\downarrow_G=t\theta\!\downarrow_G.
\)

Hence,
\(
s\theta >_G t\theta.
\)
    \item \(f = g\)
    \[
\frac{\{\, s \D t \}_b\, \cup \Delta; \;C}{\{ \; s_i \D t_i\;\}_0,\; \{\, s \D t_{i+1}, \; \dots, \; s \D t_m\;\}_1\, \cup \Delta;  \;C \; \;\cup\; gc( s> t,b)}
\]

We assume that \(s_1 =_G t_1,\;\dots,\;s_{i-1} =_G t_{i-1}\), \(s_i\theta\!\downarrow_G > t_i\theta\!\downarrow_G\),  \(s\theta\!\downarrow_G > t_{i+1}\theta\!\downarrow_G\) and \(s\theta\!\downarrow_G > t_m\theta\!\downarrow_G\).

\(
f(s_1, \!\dots,\! s_n)\theta\!\downarrow_G=f(s_1\theta\!\downarrow_G,\!\dots,\!s_n\theta\!\downarrow_G)\!>\!f(t_1\theta\!\downarrow_G,\!\dots,\!t_m\theta\!\downarrow_G)=t\theta\!\downarrow_G.
\)

    \item \(f \prec g\)
    \[
\frac{\{\, s \D t \}_b\, \cup \Delta; \;C}{\{\,s_i \D t \}_0\, \cup \Delta;  \;C \; \cup\; gc( s> t,b)}
\]

We assume that \(t = g(t_1,\dots,t_m)\).

From the assumption:
\(
s_i\!\downarrow_G > t\theta\!\downarrow_G,
\)
\(s\theta\!\downarrow
= f(s_1\!\downarrow_G,\dots,s_n\!\downarrow_G)
>
t\theta\!\downarrow_G.
\)

    \item \(t \sqsubset s\) 

 \[
\frac{\{\, s \D t \}_b\, \cup \Delta; \;C}{ \Delta;  \;C \; \cup\; gc( s> t,b)}
\]

By the nontheory subterm property, we immediately get:
\(
s\theta >_G t\theta.
\)
\end{enumerate}
   
\item Suppose \(
s\theta \in Sub_G(G)
\). Then \(b=0\), because all initial left hand sides were not in $Sub_G(G)$.

We added the ground constraint
\(
   s\theta\!\downarrow_G \;>\; t\theta\!\downarrow_G
\)
to the constraint set $C'$.

Since the derivation ends with an acyclic constraint set $C$, this
inequality is consistent with the final precedence on constants.
Hence, the induced ordering satisfies
\(
   s\theta >_G t\theta.
\)
\end{enumerate}

\end{proof}

\section{List with Length and Append example}
Now we apply the ideas of this paper to LLA.  
In particular, $NG$ is comprised of the first six (dis)equations, and $G$ must contain Equation 7.  We will assume the following precedence on the symbols of the theory, and apply LPOGC to Equation 4 of the LLA theory.

\[
app \;\succ\; len \;\succ\; car \;\succ\; cdr \;\succ\; cons \;\succ\; nil \;\succ\; s \;\succ\; zero.
\]
%
%

\begin{example}
%
Let $G = \{cons(a,b) \A c, app(c,e) \A cons(a,app(b,e))
\}$.
%
The congruence classes of $G$, are $\{\{a\},\{b\},
\{cons(a,b),c\}, \{app(b,e),c_3\}, \{cons(a,c_3),app(c,e),c_4\}
\}$.
%
%
%



\[
\begin{array}{c}

\{(app(cons(X,Y),Z) \M cons(X,app(Y,Z)),1)\}; \emptyset \\[-1pt]

\makebox[13cm]{\hrulefill\ \text{\small lpog}} \\[-1pt]

\{(app(cons(X,Y),Z) \M X,1),\;
(app(cons(X,Y),Z)\M app(Y,Z),1)\}; \emptyset \\[-1pt]

\makebox[13cm]{\hrulefill\ \text{\small lpos}} \\[-1pt]

\{(app(cons(X,Y),Z)\M app(Y,Z),1)\}; \emptyset \\[-1pt]

\makebox[13cm]{\hrulefill\ \text{\small lpoe}} \\[-1pt]

\{(cons(X,Y)\M Y,0),\;
(app(cons(X,Y),Z)\M Z,1)\}; \emptyset \\[-1pt]

\makebox[13cm]{\hrulefill\ \text{\small lpos}} \\[-1pt]

\{(app(cons(X,Y),Z)\M Z,1)\};\; c>b,\; c_4>c_3 \\[-1pt]

\makebox[13cm]{\hrulefill\ \text{\small lpos}} \\[-1pt]

\emptyset;\; c>b,\; c_4>c_3

\end{array}
\]


\end{example}

If, instead we had $G = \{cons(a,b) \A c, cons(d,c) \A b
\}$, then the inferences above would be the same, up until the last step, and $\{b > c, c > b\}$ would have been created, which is a cycle, and $G$ represents an infinite list (a list that is a strict sublist of itself).
In general, in the LLA example, if $G$ does not represent any infinite lists, then there is a derivation that will result in $\emptyset ; C$, where $C$ does not contain a cycle.
To show this, we just need to see how LPOGC will process the non-ground equations of LLA.  In Equations 1, 2 and 5, the RHS is a subterm of the LHS.  So those equations are immediately solved, with no constraints generated.  As we saw in the previous example, Equation 4 generates constraints of the form $c > d$ for all constants $c$ and $d$ where $c =_G cons(a,d)$ for some $a$.  Equation 3 does exactly the same thing.  If there is a cycle $c_0 > c_1, \cdots, c_{n-1} > c_n$ with $c_0 = c_n$ in the constraints, then this $G$ represents an infinite list.

\begin{theoremrep}
Suppose $NG$ consists of the 
equations of LLA,
and $LPOGC$ is run on $\{(s \D t, 1) \,\,|\,\, s \A t \in NG\}$;$\emptyset$.  If $G$ does not represent any infinite lists, then there is an LPOGC derivation resulting in $\emptyset ; C$, where $C$ does not contain a cycle. 

\end{theoremrep}

\begin{proof}
In Equations 1, 2 and 5, the RHS is a subterm of the LHS, so LPOGC will immediately succeed with no constraints generated.  As we saw in the previous example, Equation 4 will create a constraint $c > d$ for all ground constants $c$ and $d$ where $c =_G cons(a,d)$ for some $a$.  Equation 3 will generate those same constraints.  If there is a cycle $c_0 > c_1, \cdots, c_{n-1} > c_n$ with $c_0 = c_n$ in the constraints, then this represents an infinite list represented by $G$.
\end{proof}


Given an $NG$ containing the
equations of LLA, and a satisfiable $G$ where LPOGC has a derivation that does not produce a cycle, 
then we know that equations are oriented properly according to $<_G$. Assume $NG$ is saturated by all the rules except Superposition.
We want to show that $NG$ is then Sup($G$) saturated.  In other words, no Superposition
inference
needed for $NG$. 
But we need to examine the one potential Superposition, which is between Equations 4 and 5, when $cons(X,Y) =_G nil$.  That would require that $cons(a,b) =_G nil$ for some ground $a$ and $b$.  As we can see from Example~\ref{supg} that would require that $G$ is unsatisfiable, a contradiction. 


\begin{theoremrep}
Let $NG$ be the
equations of LLA.
Let \(\gc\)
be a set of ground (dis)equations that includes 
\({len}({nil}) \A {zero}\), where $G \models \gc$.  
If $NG$ is 
 Sup($G$)
saturated by all the rules except Superposition, 
and \(\gc\)
is satisfiable, and there is an LPOGC derivation that does not result in a cycle,
then $NG$ is 
Sup($G$)
saturated.
\end{theoremrep}

\begin{proof}
We need to show that $NG$ is 
Sup($G$) 
saturated.  

The only overlap is between Equation 4 and Equation 5. The only superposition  between Equation 4 and Equation 5 is if there are ground terms a and b such that
\(
  cons(a,b) \A _G nil.
\)
    
Since $\gc$ 
contains $len(nil)\A zero$, we would also have 
$G
\models len(cons(a,b))\A zero$. Instantiating Equation 3 yields
\(
  len(cons(a,b))\A s(len(b)),
\)
and instantiating Disequation 6 yields
\(
  s(len(b))\B zero.
\)
Combining these gives $zero\neq _Gzero$, a contradiction.  Hence, if the set of equations is saturated and satisfiable, $G$ cannot imply $cons(a,b)\A _Gnil$ for any $a,b$, and so there is no superposition between 4 and 5.
\end{proof}

\section{Related Work and Conclusion}
%


The paper \cite{DBLP:journals/jar/DrossCKP16}
discusses how a good selection of triggers will give a decision procedure.  Their approach is somewhat different from ours.  The user needs to supply a correctness and termination proof that the trigger choice will give a decision procedure.  Our method is automatic, and triggers are entire terms.
Good trigger selection is discussed from a practical point of view in \cite{DBLP:conf/cav/LeinoP16,10.1145/1670412.1670416}.   
Other papers suggest other approaches to quantifiers instead of triggers.  One successful approach is Model-Based Quantifier Instantiation~\cite{DBLP:conf/cav/GeM09}.  Several other approaches have been proposed and implemented~\cite{DBLP:conf/lpar/Rummer12,DBLP:conf/tacas/BarbosaFR17,DBLP:conf/tacas/ReynoldsBF18,DBLP:conf/frocos/FontaineS21,DBLP:conf/fmcad/ReynoldsTM14,DBLP:journals/tplp/ReynoldsTB17,DBLP:conf/tacas/NiemetzPRBT21,DBLP:conf/vmcai/HoenickeS21}.  
Our paper only  deals with first order theories with 
equality and uninterpreted function symbols,
whereas the above mentioned papers, except for \cite{DBLP:conf/tacas/BarbosaFR17}, consider other SMT theories.

We have presented an inference procedure for equational unification modulo a non-ground theory and a dynamic ground theory.  There are four new ideas to this paper.  First, we define an ordering $<_G$ that compares terms based on their normal form in the ground theory.  This ordering satisfies a weak form of monotonicity and subterm property. Second, we present the Sup($G$) inference system, which contains ideas from  E-unification modulo a ground equational theory, and the Instantiation procedure used in SMT solvers.  As opposed to other E-unification procedures, we instantiate non-ground equations to update the equational theory.  As opposed to SMT solvers, we only need instantiation at the top of the left hand side of an equation.  Most importantly, the Sup($G$) procedure is refutationally complete.  
Our last idea is the LPOGC procedure or instantiating constraints, which can allow natural orientations when it does not create a cycle, like in finite data structures.  We apply our ideas to the theory of lists with length and append.  

This work can be extended in many ways.  We plan to extend this from single equations to Horn clauses or general clauses.  It would also be very useful to   
allow quantified variables from specialized theories.  We also need to understand better what equational theories have a finite saturation.  Perhaps constructor theories are an appropriate candidate.

\bibliographystyle{plain}
\bibliography{references}
%





\appendix

\end{document}